\documentstyle[aps,preprint]{revtex}
\begin{document}
\begin{center}
\title{Non-Scissors-Mode Behaviour of Isovector Magnetic Dipole Orbital 
Transitions Involving Isospin Transfer}
\author{S. Aroua$^1$, L. Zamick$^2$, Y.Y. Sharon$^2$, E. Moya de Guerra$^3$, M.S. Fayache$^1$, 
P. Sarriguren$^3$ and A.A. Raduta$^4$}
\address{(1) D\'{e}partement de Physique, Facult\'{e} des Sciences de Tunis\\
Tunis 1060, Tunisia\\
(2) Department of Physics and Astronomy, Rutgers University\\
Piscataway, New Jersey 08855\\
(3) Instituto de Estructura de la Materia, Consejo Superior de\\
Investigaciones Cient\'{\i}ficas,\\
Serrano 119, 28006 Madrid, Spain\\
(4) Department of Theoretical Physics and Mathematics, Bucharest University, 
P.O. Box MG11,
and Institute of Physics and Nuclear Engineering, Bucharest, P.O. Box MG6, 
Romania\\}

\date{\today}
\maketitle

\begin{abstract}
We study the response of isovector orbital magnetic dipole (IOMD) transitions to 
the quadrupole-quadrupole ($Q \cdot Q$) interaction, to the isospin-conserving pairing 
interaction (ICP) and to combinations of both. We find qualitatively different behaviours 
for transitions in which the final isospin differs from the initial isospin versus 
cases where the two isospins are the same. For $N=Z$ even-even nuclei with $J^{\pi}=0^+,~T=0$ 
ground states such as $^8Be$ and $^{20}Ne$, the summed $T=0~\rightarrow~T=1$ IOMD from the ground 
state to all the $J=1,~T=1$ states in the $0 \hbar \omega$ space does not 
vanish when the $Q \cdot Q$ interaction is turned off. The pairing interaction (ICP) 
alone leads to a finite transition rate. For nuclei with $J=0^+,~T=1$ ground states such as 
$^{10}Be$ and $^{22}Ne$, the summed $T=1~\rightarrow~T=1$ IOMD $does$ vanish when the $Q \cdot Q$ 
interaction is turned off, as is expected in a good scissors-mode behaviour. However 
this is not the case for the corresponding sum of the $T=1~\rightarrow~T=2$ IOMD transitions. 
In $^{22}Ne$ (but not in $^{10}Be$) the sum of the $T=1~\rightarrow~T=2$ IOMD transitions is 
remarkably insensitive to the strengths of both 
the $Q \cdot Q$ and the ICP interactions. In $^{22}Ne$ an energy weighted-sum is similarly 
insensitive. All our calculations were carried out in the $0 \hbar \omega$ space.  
\end{abstract}
\end{center}

\section{Introduction}
The scissors mode excitation in an even-even nucleus with a $J=0^+$ ground
state is a state with quantum numbers $J=1^+$ arising from the operator
$\vec{L}_\pi -\vec{L}_\nu$ acting on the ground state. Here $\vec{L}$ is the
orbital angular momentum and $\vec{L}_\pi -\vec{L}_\nu$ is the isovector orbital angular 
momentum operator. Of course one can also get excitations from the isovector spin-operator 
$\vec{S_\pi} -\vec{S_\nu}$ as well, but in this work we focus mainly on orbital excitations.

The collective aspect of this excitation mode was emphasized as early as 1978 by N. Lo Iudice and
F. Palumbo \cite{palumbo}, who pictured the scissors mode as a vibration of the deformed 
symmetry axis of all the protons against that of all the neutrons. Experimental evidence for the 
existence of a scissors mode in $^{156}Gd$ was obtained by D. Bohle $et~al.$ \cite{Bohle} of 
the Darmstadt group in 1984. It was soon realized, however, that the initial picture was too 
extreme and that mainly only the valence nucleons contributed to the scissors mode excitation, 
because one cannot have $M1$ excitations of any kind in a closed major shell. The $IBA$ 
treatment of Iachello \cite{iachel} does involve only valence bosons and gives more reasonable 
results. Shell model approaches also involve valence nucleons, see for instance the early work 
of Zamick \cite{zamick} and that of Chaves and Poves \cite{chaves}.

It was realized that one could connect the isovector orbital strength $B(M1)$ to the 
nuclear deformation parameter $\delta$ or, alternatively, to the electric quadrupole strength 
$B(E2)$ in a given nucleus. For example, Rohoziwski and Greiner \cite{Roho} have obtained an 
expression in which the $B(M1)$ scissors-mode strength is proportional to $\delta ^2$. In 
another approach known as the ``Generalized Coherent State Model'', Raduta $et~al.$ have also 
obtained the same $\delta ^2$ dependence of the $M1$ strength \cite{raduta}, 
see also the works of Hamamoto and Magnusson \cite{hama} and Enders $et~al.$ \cite{Richter}.

The connection with deformation has led to the suggestion of using a quadrupole-quadrupole 
interaction in shell model calculations which aim to study the scissors mode. In a previous work 
by one of the present authors \cite{zz}, it was shown that with a simple $Q\cdot Q$ interaction 
the summed energy-weighted isovector orbital $M1$ strength was proportional not simply to the 
$B(E2)$ but rather to $\sum B(E2, isoscalar)-\sum B(E2, isovector)$ where the first term was 
calculated with the effective nuclear charges $e_p = 1$, $e_n = 1$ and the second term with 
$e_p = 1$ and $e_n = -1$. This expression has the correct behaviour when one of the major shells 
of neutrons or protons is closed. In that case, one can have finite $B(E2)$'s but vanishing 
$B(M1$'s). Indeed, for such a case, the isoscalar and isovector $B(E2)$ sums are the same and 
their difference vanishes.

There have also been related calculations which included excitations to higher shells 
\cite{moya1,smith,Iudice}. In Ref. \cite{moya1} it was noted that in the simple Nilsson model 
the amount of energy-weighted strength at $2 \hbar \omega$ is the same as at $0 \hbar \omega$. 
More detailed calculations with an isovector $Q \cdot Q$ interaction have also been 
performed \cite{hama}.

In this work we shall consider a combination of a $ Q\cdot Q$ interaction 
and an isospin-conserving pairing interaction. We shall examine light nuclei with a focus on the 
difference in behaviour of the transitions in which the final isospin is
different from the initial isospin versus transitions in which the isospin is unchanged. 
In heavy nuclei with large neutron excess, such an analysis is not possible at present. 

We would argue that if indeed the $B(M1)$ is simply connected to deformation (as is the case for 
the scissors states), then the stronger we make the $Q \cdot Q$ strength relative to the pairing 
strength the stronger should be the summed isovector orbital $B(M1)$. We will see in the next 
sections whether this is what happens in various cases.

One word about the isospin-conserving pairing interaction (ICP). If one limits
oneself to a pairing interaction involving particles of one kind, then the double commutator 
of the isovector orbital angular momentum operator with such a (non-isospin conserving) pairing
interaction will vanish. So if this were the only interaction present, the $B(M1)$ would vanish. 
But when we include in the pairing interaction also the $J=0$ pairing between a neutron and a 
proton, the situation becomes more complicated as will be seen in the following sections.

\section{An Overview of the Calculations}

The two-body interaction that we use in our calculations can be written as the
sum of the $Q\cdot Q$ term and the Isospin-Conserving Pairing Interaction
($ICP$) term. The ($ICP$) term is defined by:

\[{\langle j_1 j_2 | V_{pairing} | j_3 j_4 \rangle}^{J,T} =-\frac{G}{\sqrt{1+
\delta_{1,2}}\sqrt{1+\delta_{3,4}}} \delta_{J,0}\delta_{T,1}
\sqrt{(2j_1+1)(2j_3+1)}\delta_{j_1,j_2}\delta_{j_3,j_4}\]

\noindent whereas the $Q \cdot Q$ interaction is $-\chi \sum_{i < j}Q(i) \cdot Q(j) = -\chi 
\sum_{i < j}r_i^2Y_2(i) \cdot r_j^2Y_2(j)$.

\noindent Here $G$ and $\chi$ are the respective strengths of these two interactions.

We carry out calculations with a hamiltonian that involves a kinetic energy
and harmonic oscillator single-particle terms plus the $Q \cdot Q$
and $ICP$ two-body interaction terms. These calculations are carried out in
the $0 \hbar \omega $ space using the computer code $OXBASH$ \cite{oxbash}.

We consider the four nuclei $^8Be$, $^{10}Be$, $^{20}Ne$ and $^{22}Ne$. For
each nucleus we calculate the summed isovector orbital transition strength
$B(M1)$ from the ground state to all the $J=1^+$ states with a given $T$ in our
$0 \hbar \omega $space. For the $N=Z$ nuclei $^8Be$ and $^{20}Ne$, we sum the
$T=0~\rightarrow~ T=1$ isovector orbital $M1$ excitations from the $J=0^+$,
$T=0$ ground state to all the $J=1^+$, $T=1$ states in the $0 \hbar \omega$ 
space. For $^{10}Be$ and $^{22}Ne$, we can have both $T=1~\rightarrow~ T=1$
and $T=1~\rightarrow~ T=2$ transitions from the $J=0^+,~T=1$ ground state
to $1^+$ states, and we sum their strengths separately.

%\begin{table}
%\caption{Values of isovector orbital summed strenghs when $G=0$ and when $\chi=0$}
%\begin{tabular}{|c|ccc|}
%\hline
%  Nucleus   &          Mode          &   $G=0$ & $\chi=0$ \\
%\hline
%   $^8Be$   & $T=0~\rightarrow~ T=1$ &  2.547  &  0.728   \\
%            &                        &         &          \\
%   $^{10}Be$& $T=1~\rightarrow~T=1$  &  0.358  &  0.000   \\
%            & $T=1~\rightarrow~T=2$  &  0.596  &  0.218   \\
%            &                        &         &          \\
%   $^{20}Ne$& $T=0~\rightarrow~T=1$  &  4.673  &  2.447   \\
%            &                        &         &          \\
%   $^{22}Ne$& $T=1~\rightarrow~T=1$  &  3.659  &  0.000   \\
%            & $T=1~\rightarrow~T=2$  &  1.098  &  1.175   \\
%\hline
%\end{tabular}
%\end{table}

In table I we give the values of the summed isovector orbital $M1$ strengths
in the extreme limits of a pure $Q \cdot Q$ ($G=0$) and a pure ICP ($\chi = 0$) 
interaction separately.

Note that, in Figs. 1-4, the $G=0$ results are those at the extreme left of
each figure while the $\chi = 0$ values are the asymptotic values of the
curves at the extreme right of each figure and beyond.

\section{Results of the Calculations}

\subsection{A Brief Discussion of Isoscalar Transitions}

For all combinations of the different strenghts of the $Q \cdot Q$ and ICP interactions, the 
summed isoscalar orbital transitions vanish in all nuclei. This is true for 
$T=0~\rightarrow~T=0$ and $T=0~\rightarrow~T=1$ transitions in $^8Be$ and $^{20}Ne$, and it is 
also true for $T=1~\rightarrow~T=1$ and $T=1~\rightarrow~T=2$ transitions in $^{10}Be$ and 
$^{22}Ne$.

\subsection{Isovector Transitions in $^8Be$ and $^{20}Ne$ with $T=0$ ground states}

The $N=Z$ nuclei $^8Be$ and $^{20}Ne$ have $T=0$ ground-state isospins, and isovector 
transitions from such a state will therefore lead only to $T=1$ excited states. In Figs. 1 and 
2, we show the results of calculations of the isovector orbital summed $M1$ strengths 
as a function of the parameter $R=G/18\bar{\chi}$ where $\bar{\chi}=\frac{5b^4}{32\pi}\chi$. 
In the $SU(3)$ model, the excitation energy of the $2^+_1$ state is $E_x(2^+_1)=18\bar{\chi}$. 
Hence the ratio $G/18\bar{\chi}$ is a dimensionless parameter which tells us how the strength 
of the ICP interaction relates to an $SU(3)$-model estimate of $E_x(2^+_1)$. The limiting case 
where the $Q \cdot Q$ interaction is turned off corresponds to $R=G/18\bar{\chi}$ going to 
infinity, while for the pure $Q \cdot Q$ case $R=G/18\bar{\chi}$ is zero. 
Essentially, our parameter $R=G/18\bar{\chi}$ is proportional to the ratio
of the relative strengths of the $ICP$ and $Q \cdot Q$ interactions.

Examining Figs. 1 and 2 for $^8Be$ and $^{20}Ne$, respectively, we see that as
we move rightwards from $R=0$ the summed isovector orbital magnetic dipole strength starts to 
go down approximately linearly from an initially large value (for $R=0$). 
But then this summed $B(M1)$  levels off and reaches a $non-zero$ asymptotic value. This tells us 
that as we add an ICP interaction to what was originally a pure $Q \cdot Q$ interaction the 
summed isovector orbital magnetic dipole strength decreases. At the other extreme, for large $R$ 
we get a non-zero value of this $M1$ strength. In other words, if we just have 
an ICP interaction we get a finite summed isovector orbital strength from $T=0$ to $T=1$. This is 
a non-scissors-mode contribution and as such is a new mode of excitation. However, the 
addition of a $Q \cdot Q$ interaction to an $ICP$ interaction does lead to an enhancement of the 
isovector orbital strength, and a scissors-mode contribution is present in that case as well. 

\subsection{Isovector Transitions in Nuclei with $T=1$ Ground States, $^{10}Be$
and $^{22}Ne$}

For nuclei with $T=1$ ground states, such as $^{10}Be$ and $^{22}Ne$, we consider 
separately the summed isovector orbital $T=1~\rightarrow~T=1$ transitions and 
$T=1~\rightarrow~T=2$ transitions, see Figs. 3 and 4. 

For the $T=1$ to $T=1$ transitions, the summed isovector orbital magnetic dipole 
strength vanishes in the limiting case of a vanishing $Q \cdot Q$ interaction 
(corresponding to $R$ going to infinity). This vanishing for a pure ICP interaction 
is shown analytically in Appendix A. This $T=1~\rightarrow~T=1$ mode 
displays the simple scissors-mode behaviour; for a fixed strength $\chi$ of the 
$Q \cdot Q$ interaction, the summed isovector orbital magnetic dipole strength 
decreases as we increase the ICP strength $G$. Thus, the pairing interaction 
serves to decrease the summed isovector orbital strength. 

For $T=1~\rightarrow~T=2$ transitions the behaviour is quite different. In $^{10}Be$, 
as we increase the parameter $R$, the summed isovector orbital magnetic 
dipole strength decreases rapidly from a large value, but reaches an asymptotic value 
for large $R$ that is non-zero (see Fig. 3). 

For $^{22}Ne$, we find that the summed isovector orbital magnetic dipole strength 
hardly changes at all as we increase the value of the parameter $R$, and the value of the summed 
strength always remains very close to unity (see Fig. 4). More specifically the strength is 
1.098 for $R = 0 $, decreases to 0.985 for $R \simeq 0.4 $ and then increases to 1.174 for 
$R \rightarrow \infty$. Thus the summed transition  
strength is rather insensitive to what we choose as the strength parameters of the 
$Q \cdot Q$ and ICP interactions, provided of course that we do not change their 
overall signs. We can call these newly studied isovector orbital $M1$ excitations of the 
$J^{\pi}=1^+~T=2$ states in $^{22}Ne$ the $insensitive$ modes.

\subsection{The energy-weighted $B(M1)$ sums in $^{22}Ne$}

In $^{22}Ne$, it is interesting to study also two energy-weighted isovector
orbital $B(M1)$ sums. If we consider an initial state $i$ and many final states $f$,
the energy-weighted $B(M1)$ sum ($EWS$) is then defined as 
$EWS \equiv \sum_f (E_i-E_f)B(M1:i \rightarrow f)$.

One energy-weighted isovector orbital $B(M1)$ sum is 
$A \equiv (EWS)_{T=1\rightarrow T=1}$, which considers all the isovector orbital $M1$ 
transitions from the $0^+_1$, $T=1$ ground state to all the $1^+$, $T=1$ states in our 
$0 \hbar \omega$ model space.
The other energy-weighted sum is $B \equiv (EWS)_{T=1\rightarrow T=2}$ which considers all
the isovector orbital $M1$ transitions from the $0^+_1,~T=1$ ground state to all the $1^+$, 
$T=2$ states in our model space. For a given value of $\bar{\chi}$, we studied the behaviour of 
the quantities $A$, $B$, and $A+B$ as a function of our usual parameter 
$R= \frac{G}{18 \bar{\chi}}$. The results are plotted in Fig. 5 for $R$ ranging from 0 to 1. 
Three curves are drawn there corresponding respectively to $A$, $B$ and $A+B$.

From Fig. 5, it is clear that for $0<R<1$ the sum $A+B$ is insensitive to $R$.
This sum varies in this region only by $13\%$ although A and B individually 
decrease respectively, by close to a factor of three in each case. Note that in Fig. 5 we 
used a single value of $\bar{\chi}$ (with $\bar{\chi} \simeq 0.1$), and we varied $R$ by 
simply varying $G$. However, we have found that the quantity $(A+B)/\bar{\chi}$ still remains 
insensitive to $R$ as we vary both $G$ and $\bar{\chi}$ independently.

\section{Conclusion}

In this paper we have reported on the results of a comparative shell-model study of Isovector 
Orbital Magnetic Dipole Transitions (IOMD) in light even-even nuclei using a combination 
of a quadrupole-quadrupole ($Q\cdot Q$) interaction and an isospin-conserving pairing (ICP) 
interaction. We have found that the IOMD transitions involving $\Delta T =1$ states have a 
non-scissors mode behaviour in the sense that the IOMD strength remains finite even for a 
vanishing deformation-inducing interaction such as $Q\cdot Q$. Comparing the results for the 
extreme situations of $\chi=0$ and $G=0$ (absence of the $Q\cdot Q$ interaction, and absence 
of the ICP interaction, respectively), one notices that the contribution of the $Q\cdot Q$ 
interaction prevails over that of the ICP one. An exception is the case of $^{22}Ne$ where the 
contribution of ICP is slightly larger than that of $Q\cdot Q$. Since this case reveals quite 
unexpected properties, we refer to the corresponding mode as a {\em new mode}. In this mode, 
the isovector orbital $M1$ excitation from $T=1$ to $T=2$ is relatively insensitive to the 
strength of the $Q \cdot Q$ and/or that of the pairing interaction. Furthermore, the summed 
energy-weighted isovector orbital $B(M1)$, when added for both $T=1 \rightarrow T=1$ and $T=1 
\rightarrow T=2$ is again quite insensitive to the ratio of the isospin-conserving pairing 
interaction strength to the quadrupole-quadrupole interaction strength.

\section{Acknowledgements}
This work was supported by NATO Linkage Grant PST 9781 58, by U. S. Dept of
Energy grant DE-FG02-95ER-40940, MCYT (Spain) Contract numbers PB98-0676 and 
BFM 2002-0356, and a Stockton College Summer Faculty Research Fellowship Grant.

\pagebreak

\newpage
\section{APPENDIX A}

In this Appendix we show that in nuclei like $^{10}Be$ and $^{22}Ne$, with a pure $T=1$, 
isospin-conserving pairing interaction, the isovector orbital $B(M1)$ excitation connects the 
0$^+$, $T=1$ ground state only to states with $L=1$, $S=0$, $T=2$.

For such an interaction the ground state for the $4\nu \oplus 2\pi $ nuclei in a given 
(degenerate) major $N-$shell (i.e., $N=1$ for $^{10}$Be, $N=2$ for $^{22}$Ne) will be of 
the general form

\begin{equation}
\left| GS \right\rangle = A \left\{ {\cal S}(\pi \pi)^1_0 \otimes 
(\nu \nu)^1_0 \otimes(\nu \nu)^1_0 \right\} ^1_0 +
B \left\{ {\cal S}(\nu \pi)^1_0 \otimes 
(\nu \pi)^1_0 \otimes(\nu \nu)^1_0 \right\} ^1_0
\end{equation}
where the upper labels stand for isospin and the lower labels are for angular
momentum ($j=0,\ell=0,s=0,$ for each pair in the ground state). ${\cal S}$
indicates symmetrized product of pairs. 

To use a shorter notation (since they are all $s=0,\ell=0$ pairs), we write only
the $t,t_z$ of the pairs. We write

\begin{equation}
(\pi \pi )= {\cal A}^{1-1};\quad 
(\nu \nu )= {\cal A}^{11};\quad
(\pi \nu )^1_0= {\cal A}^{10};\quad 
(\pi \nu )^0_1= \tilde{\cal A}^{00}
\end{equation}
where the super index indicates isospin labels (${\cal A}^{t,t_z}$).

It is easy to show that $\vec{L}_\pi -\vec{L}_\nu$ (and likewise 
$\vec{S}_\pi -\vec{S}_\nu$ and  $\vec{J}_\pi -\vec{J}_\nu$) acting on the first
component  ($\propto A $) gives zero because

\begin{eqnarray}
\left( \vec{L}_\pi -\vec{L}_\nu \right) \left| \left\{ {\cal S} \left( 
{\cal A}^{1-1}\otimes {\cal A}^{11}\otimes {\cal A}^{11} \right) ^{11}_{00}
\right\} \right\rangle &=& 
\left| \left\{ {\cal S} \left[ \left( \left( \vec{\ell}^{(1)}_\pi + 
\vec{\ell}^{(2)}_\pi \right) {\cal A}^{1-1}\right) \otimes {\cal A}^{11}
\otimes {\cal A}^{11} \right] \right\} ^{T1}_{00} \right\rangle  \nonumber \\
&-& \left| {\cal S} \left[ {\cal A}^{1-1} \otimes
\left( \left( \vec{\ell}^{(1)}_\nu + 
\vec{\ell}^{(2)}_\nu \right) {\cal A}^{11}\right) \otimes {\cal A}^{11}
\right] \right\rangle + \cdots
\end{eqnarray}
and

\begin{eqnarray}
&&\left( \vec{\ell}^{(1)}_\pi + 
\vec{\ell}^{(2)}_\pi \right) {\cal A}^{1-1} =0 \\
&&\left( \vec{\ell}^{(1)}_\nu + 
\vec{\ell}^{(2)}_\nu \right) {\cal A}^{11} =0
\end{eqnarray}
i.e., $(\pi\pi)^1_0$ is an eigenstate of 
$\vec{\ell}^{(1)}_\pi + \vec{\ell}^{(2)}_\pi \equiv \vec{\ell}_{12}$ with eigenvalue 
$\ell _{12}=0$.

Similarly it is a simple matter to show that 
$\vec{\ell}^{(1)}_\pi - \vec{\ell}^{(2)}_\nu$ acting on a $\pi\nu$ pair with
$T=1,L=0,S=0$ transforms this pair into one with $T=0,L=1,S=0$
(likewise $\vec{S}^{(1)}_\pi - \vec{S}^{(2)}_\pi$ transforms it into one with $T=0,L=0,S=1)$. 
Therefore we can write that 
$(\vec{\ell}^{(1)}_\pi - \vec{\ell}^{(2)}_\nu ){\cal A}^{10}=C \tilde{\cal A}^{\
00}$,
where $\tilde{\cal A}^{00}$ has $L=1,S=0$, and that

\begin{equation}
\left( \vec{L}_\pi -\vec{L}_\nu \right) \left| GS \right\rangle =
B\left( \vec{L}_\pi -\vec{L}_\nu \right) \left\{ \left[ {\cal S} \left(
{\cal A}^{10}\otimes{\cal A}^{10}\otimes{\cal A}^{11}\right) \right] ^{11}_{00}\
\right\}
\end{equation}
Let us first see what the term  
$\left[ {\cal S} \left( {\cal A}^{10}\otimes{\cal A}^{10}\otimes{\cal A}^{11}\right) \right] ^1_0$ looks like:

\begin{eqnarray}
{\cal S} \left( {\cal A}^{10}\otimes{\cal A}^{10}\otimes{\cal A}^{11}\right) 
&=& \frac{1}{{\cal N}} \left\{  {\cal A}^{10}\otimes{\cal A}^{10}\otimes{\cal A\
}^{11}
+  {\cal A}^{10}\otimes{\cal A}^{11}\otimes{\cal A}^{10}+
{\cal A}^{11}\otimes{\cal A}^{10}\otimes{\cal A}^{10} \right\} ^{11}_{00} \nonumber \\
&=&\frac{1}{{2 \cal N}} \{ \sum_T \left\{ {\cal A}^{10}\otimes \left[  {\cal A}^{10}\
\otimes
{\cal A}^{11} \right] ^{T1}\right\}^{11} +
\left\{ \left[ {\cal A}^{10}\otimes {\cal A}^{10}\right]^{T0} \otimes
{\cal A}^{11} \right\}^{11} \nonumber \\
&+&  \left\{ \left[ {\cal A}^{10}\otimes {\cal A}^{11} \right] ^{T1}\otimes
{\cal A}^{10} \right\}^{11} +
\left\{ {\cal A}^{10}\otimes \left[ {\cal A}^{11} \otimes
{\cal A}^{10}\right] ^{T1} \right\}^{11} \nonumber \\
&+&  \left\{ {\cal A}^{11}\otimes \left[ {\cal A}^{10} \otimes
{\cal A}^{10} \right] ^{T0}\right\}^{11} +
\left\{ \left[ {\cal A}^{11}\otimes {\cal A}^{10}\right] ^{T1} \otimes
{\cal A}^{10}\right\}^{11} \}
\end{eqnarray}
Now we use

\begin{eqnarray}
&&\sum_T \left[ {\cal A}^{10}\otimes {\cal A}^{11} \right] ^{T1} = \sum_{T=1,2}\
\left\langle 1011|T1\right\rangle {\cal A}^{10}{\cal A}^{11} \nonumber \\
&&\sum_T \left[ {\cal A}^{11}\otimes {\cal A}^{10} \right] ^{T1} = \sum_{T=1,2}\
(-1)^T\left\langle 1011|T1\right\rangle {\cal A}^{11}{\cal A}^{10} \nonumber \\\
&&\sum_T \left[ {\cal A}^{10}\otimes {\cal A}^{10} \right] ^{T0} = \sum_{T=0,2}\
\left\langle 1010|T0\right\rangle {\cal A}^{10}{\cal A}^{10} \nonumber \\
&&\left\{ {\cal A}^{10} \otimes \left[ {\cal A}^{10}\otimes {\cal A}^{11} \right]
^{T1} \right\} ^{11} = \sum _{T=1,2} \left\langle 10T1|11\right\rangle 
\left\langle 1011|T1\right\rangle  {\cal A}^{10} {\cal A}^{10} {\cal A}^{11} \\\
&&\vdots
\end{eqnarray}
to write

\begin{eqnarray}
&&\left\{ {\cal S} \left( {\cal A}^{10}\otimes {\cal A}^{10}\otimes 
{\cal A}^{11} \right) \right\} ^1_1 = \nonumber \\
&& \frac{1}{{\cal N}} \left\{ \sum_{T=1,2} \left\langle 1011|T1\right\rangle  
\left\langle 10T1|11\right\rangle \left[ {\cal A}^{10}{\cal A}^{10}{\cal A}^{11\
} 
+(-1)^T 2 {\cal A}^{10}{\cal A}^{11} {\cal A}^{10}+ {\cal A}^{11} {\cal A}^{10}\
 {\cal A}^{10} \right] \right. \nonumber \\
&&+ \left. \sum_{T'=0,2} \left\langle 1010|T'0\right\rangle  
\left\langle T'011|11\right\rangle \left[ {\cal A}^{10}{\cal A}^{10}{\cal A}^{1\
1} 
+(-1)^{T'} {\cal A}^{11}{\cal A}^{10} {\cal A}^{10}\right] \right\}
\end{eqnarray}
Now we have that

\begin{eqnarray}
\left( \vec{L}_\pi -\vec{L}_\nu \right){\cal A}^{10}{\cal A}^{10}{\cal A}^{11} \
&=&
\left[ \left( \vec{\ell}^{(1)}_\pi - \vec{\ell}^{(2)}_\nu \right){\cal A}^{10}
\right] {\cal A}^{10}{\cal A}^{11}+ {\cal A}^{10} \left[ \left( \vec{\ell}^{(1)\
}_\pi 
- \vec{\ell}^{(2)}_\nu \right) {\cal A}^{10}\right] {\cal A}^{11} \nonumber \\
&=&C\left[  \tilde{\cal A}^{00}{\cal A}^{10}{\cal A}^{11}+{\cal A}^{10}
 \tilde{\cal A}^{00}{\cal A}^{11} \right]
\end{eqnarray}
and so on for every component. Now we use the fact that

\begin{equation}
\left[  \tilde{\cal A}^{00},{\cal A}^{10} \right] =0;\quad
\left[  \tilde{\cal A}^{00},{\cal A}^{11} \right] =0
\end{equation}
and write in all that

\begin{eqnarray}
\left( \vec{L}_\pi -\vec{L}_\nu \right) \left| GS\right\rangle &=&
\frac{1}{{2 \cal N}} B C  \tilde{\cal A}^{00}\left\{ 
 \sum_{T=1,2}  \left\langle 1011|T1\right\rangle \left\langle 10T1|11\right\rangle
\right. \nonumber \\
&&\times \left[ 2{\cal A}^{10}{\cal A}^{11} + (-1)^T \left( 2 {\cal A}^{11}{\cal A}^{10}
+2{\cal A}^{10}{\cal A}^{11} \right) +2 {\cal A}^{11}{\cal A}^{10} \right]
\nonumber \\
&&\left. + \sum_{T'=0,2} \left\langle 1010|T'0\right\rangle 
\left\langle T'011|11\right\rangle \left( 2 {\cal A}^{10}{\cal A}^{11}
+2 {\cal A}^{11}{\cal A}^{10} \right) \right\} .
\end{eqnarray}
Now we construct from these the total isospin states with $T=1$ and $T=2$ (there 
cannot be $T=0$ because $T_z=1$).

Since $\tilde{\cal A}^{00}$ has $T=0$ and $L=1$, while the rest have $T=1$ and
$L=0$, we know that the total state has $L=1$ with a coupling coefficient equal to 
one and we do not need to worry about this pair. We find that

\begin{eqnarray}
\left( \vec{L}_\pi -\vec{L}_\nu \right) \left| GS\right\rangle &=&
\frac{1}{2 {\cal N}} B C \left[ 4\left\langle 1011|21\right\rangle 
\left\langle 1021|11\right\rangle +2 \left( \frac{-1}{\sqrt{3}}+
 \left\langle 1010|20\right\rangle \left\langle 2011|11\right\rangle \right) \right]
\nonumber \\
&&\times \sum_{T=1,2} \left( \left\langle 1011|T1\right\rangle +\left\langle 
1110|T1
\right\rangle \right) \left\{ {\cal S} \left(  \tilde{\cal A}^{00}\otimes {\cal A}^{10}\otimes {\cal A}^{11}\right) \right\}^{T1}_{L=1,S=0} \nonumber \\
&=&\frac{1}{{\cal N}} B C \left[ 4\left\langle 1011|21\right\rangle 
\left\langle 1021|11\right\rangle +2 \left( \frac{-1}{\sqrt{3}}+
 \left\langle 1010|20\right\rangle \left\langle 2011|11\right\rangle \right) \right]
\nonumber \\
&&\times \left\langle 1011|21\right\rangle \delta_{T2}  
\left\{ {\cal S} \left(  \tilde{\cal A}^{00}\otimes {\cal A}^{10}
\otimes {\cal A}^{11} \right) \right\}^{T1}_{L=1,S=0}
\end{eqnarray}
Then $\left( \vec{L}_\pi -\vec{L}_\nu \right) \left| GS\right\rangle$
is orthogonal to the state with $L=1, S=0, T\neq 2, T_z=1$.

\pagebreak
\begin{table}
\caption{Values of isovector orbital summed strenghs when $G=0$ and when $\chi=0$}
\begin{tabular}{|c|ccc|}
\hline
  Nucleus   &          Mode          &   $G=0$ & $\chi=0$ \\
\hline
   $^8Be$   & $T=0~\rightarrow~ T=1$ &  2.547  &  0.728   \\
            &                        &         &          \\
   $^{10}Be$& $T=1~\rightarrow~T=1$  &  0.358  &  0.000   \\
            & $T=1~\rightarrow~T=2$  &  0.596  &  0.218   \\
            &                        &         &          \\
   $^{20}Ne$& $T=0~\rightarrow~T=1$  &  4.673  &  2.447   \\
            &                        &         &          \\
   $^{22}Ne$& $T=1~\rightarrow~T=1$  &  3.659  &  0.000   \\
            & $T=1~\rightarrow~T=2$  &  1.098  &  1.175   \\
\hline
\end{tabular}
\end{table}

\pagebreak

{\bf Figure Captions}
\vspace{0.5in}

{\bf Fig. 1:} For $^8Be$, the sum of the isovector orbital $B(M1)$'s from the 
$J=0^+_1, T=0$ ground state to all the $J=1^+, T=1$ states in the $0 \hbar\omega$ space. The 
parameter $R$, defined in the text, is proportional to the ratio
of the strengths of the Isospin-Conserving Pairing and $Q \cdot Q$ interactions.

\vspace{0.5in}

{\bf Fig. 2:} For $^{20}Ne$, the sum of the isovector orbital $B(M1)$'s from the 
$J=0^+_1, T=0$ ground state to all the $J=1^+, T=1$ states in the $0 \hbar \omega$ space.

\vspace{0.5in}

{\bf Fig. 3:} For $^{10}Be$, the separate sums of the $T=1~\rightarrow~T=1$
and the $T=1~\rightarrow~T=2$ isovector orbital $B(M1)$'s from the $J=0^+_1, T=1$ ground state.

\vspace{0.5in}

{\bf Fig. 4:} for $^{22}Ne$, the separate sums of the $T=1~\rightarrow~T=1$ and the 
$T=1~\rightarrow~T=2$ isovector orbital $B(M1)$'s from the $J=0^+_1, T=1$ ground state.

\vspace{0.5in}

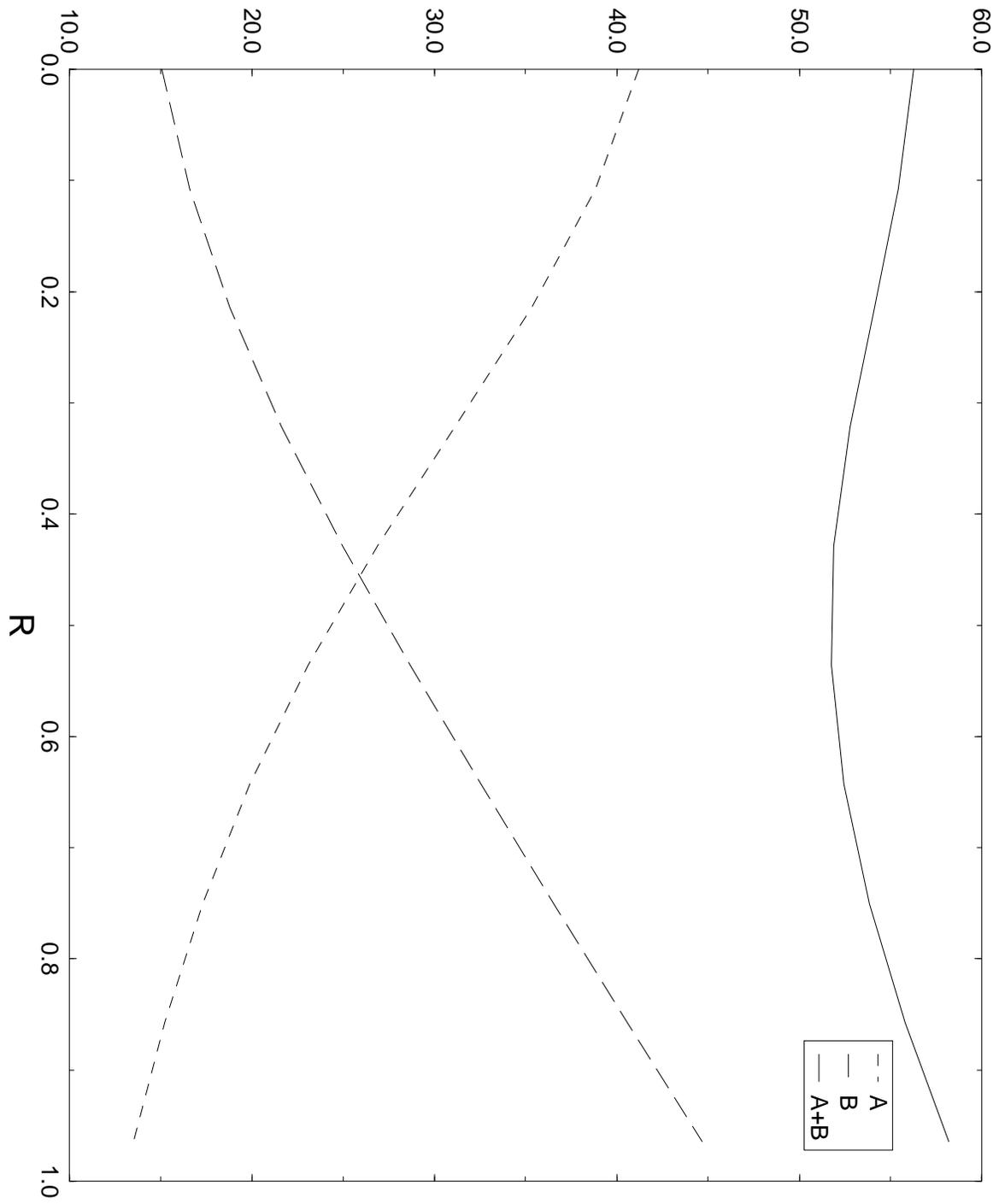
{\bf Fig. 5:} For $^{22}Ne$, the Energy-Weighted Isovector Orbital $B(M1)$ sums
$A$, $B$, and $A+B$ as defined in the text. $A+B$ is insensitive to $R$.

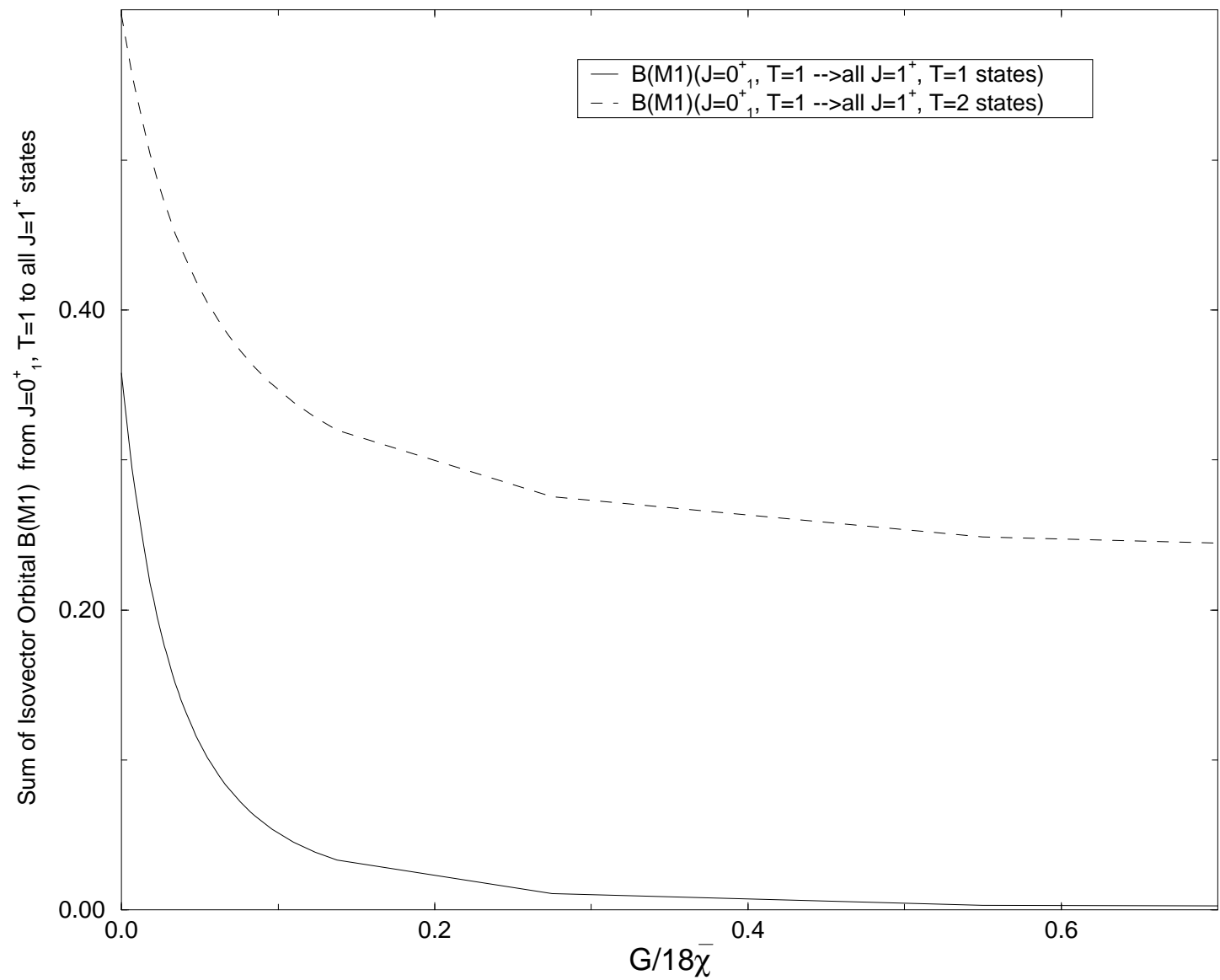
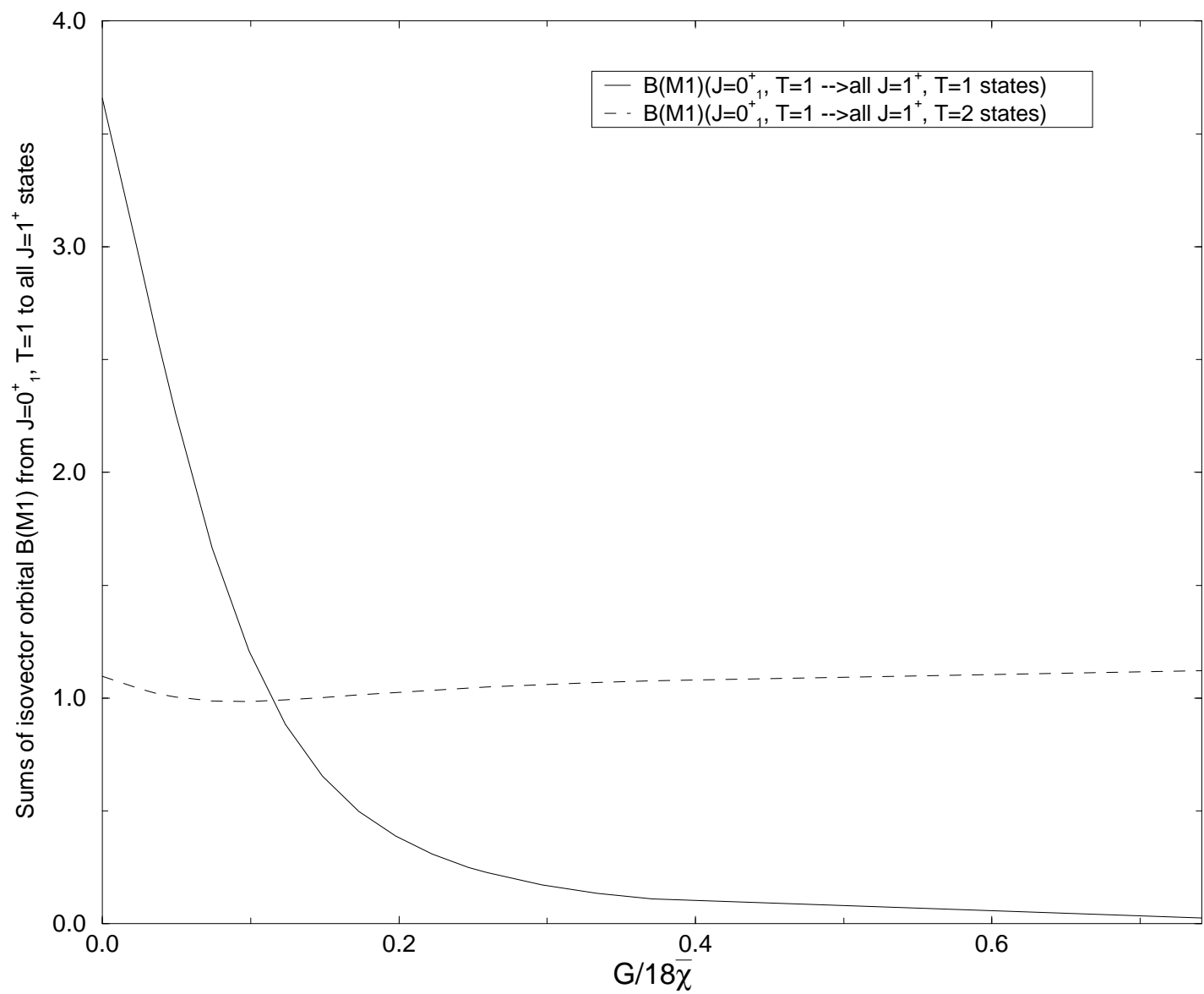
\end{document}